\documentstyle[12pt]{article}
\newlength{\extraspace}
\newlength{\extraspaces}
\newcommand{\be}{\begin{equation}
\addtolength{\abovedisplayskip}{\extraspaces}
\addtolength{\belowdisplayskip}{\extraspaces}
\addtolength{\abovedisplayshortskip}{\extraspace}
\addtolength{\belowdisplayshortskip}{\extraspace}}
\newcommand{\ee}{\end{equation}}
\newcommand{\ba}{\begin{eqnarray}
\addtolength{\abovedisplayskip}{\extraspaces}
\addtolength{\belowdisplayskip}{\extraspaces}
\addtolength{\abovedisplayshortskip}{\extraspace}
\addtolength{\belowdisplayshortskip}{\extraspace}}
\newcommand{\ea}{\end{eqnarray}}
\newcommand{\nonu}{\nonumber \\[.5mm]}
\newcommand{\A}{&\!\!\!}
\begin{document}
\thispagestyle{empty}
\begin{flushright}
SIT-LP-06/10 \\
{\tt hep-th/0611051} \\
October, 2006
\end{flushright}
\vspace{7mm}
\begin{center}
{\large{\bf On Yukawa and mass terms in nonlinear supersymmetry 
equivalent renormarizable theory}} \\[20mm]
{\sc Kazunari Shima}
\footnote{
\tt e-mail: shima@sit.ac.jp} \ 
and \ 
{\sc Motomu Tsuda}
\footnote{
\tt e-mail: tsuda@sit.ac.jp} 
\\[5mm]
{\it Laboratory of Physics, 
Saitama Institute of Technology \\
Fukaya, Saitama 369-0293, Japan} \\[20mm]
\begin{abstract}
We show explicitly in $D = 2$ that 
$N = 2$ nonlinear supersymmetric (SUSY) Volkov-Akulov model 
is equivalent to a spontaneously broken $N = 2$ linear SUSY 
interacting theory containing the ordinary Yukawa interactions and mass terms, 
which is renormalizable, by using SUSY invariant relations. 
\end{abstract}
\end{center}

\newpage

Dynamics of Nambu-Goldstone (NG) fermions \cite{SS}-\cite{O} 
which indicates a spontaneous supersymmetry breaking (SSB)
are described by the nonlinear supersymmetric (NLSUSY) model 
of Volkov-Akulov (VA) \cite{VA}. 
The linearization of NLSUSY theory gives various renormalizable 
linear (L) supermultiplet \cite{WZ1}-\cite{Fa} 
with the Fayet-Iliopoulos (FI) $D$ term which shows the SSB. 
Indeed, $N = 1$ NLSUSY VA model is recasted into 
a scalar supermultiplet \cite{IK}-\cite{UZ} 
or a U(1) axial vector one \cite{IK,STT1}. 
$N = 2$ NLSUSY VA model is (algebraically) equivalent to 
a SU(2) $\times$ U(1) vector supermultiplet \cite{STT2}. 
We have also recently shown the (algebraic) equivalence 
between $D = 2$, $N = 3$ NLSUSY VA model 
and a SO(3) vector supermultiplet, 
which contains new features common to the large $2N + 1$ SUSY 
\cite{STlin}. 

The linearization of NLSUSY gives deep insight 
into the works towards a composite unified theory 
of spacetime and matter based upon SO(10) super-Poincar\'e group, 
i.e. the superon-graviton model (SGM) \cite{KS}, 
which has been constructed as the NLSUSY general relativity (GR) \cite{KS,ST}. 
In order to obtain further informations 
about the low energy physics of NLSUSY GR, 
it is important to investigate interactions among low spin states 
in NL representation of SUSY. 
For example, as is well known, 
Yukawa-type interactions and mass terms are to be included 
in the scalar supermultiplet of Wess-Zumino \cite{WZ1}. 
However, those terms have not been discussed yet explicitly 
from the veiwpoint of NLSUSY so far to our knowledge. 

In this letter we focus on the Yukawa interactions and mass terms 
in $D = 2$ SO(2) (U(1)) vector supermultiplet for simplicity of calculations 
and study those terms in $N = 2$ NLSUSY VA model. 
\footnote
{
The Yukawa term for $N = 1$ SUSY in $D = 2$ 
vanishes by itself due to $({\rm Majorana\ NG\ fermion}\ \psi)^3 \equiv 0$. 
}
In order to do this, 
we linearize the $N = 2$ NLSUSY of VA in $D = 2$ 
by constructing SUSY (and SO(2)) invariant relations 
between NG fermions and component fields of the vector supermultiplet. 
The equivalence of $N = 2$ NLSUSY VA action 
to a free LSUSY action with a FI term 
is proved to all orders by using those relations. 
We also show explicitly that the LSUSY Yukawa interactions and mass terms 
in the vector supermultiplet vanish respectively 
by using the SUSY invariant relations 
due to cancellations among the (NG fermion) terms. 
This fact means that the $N = 2$ NLSUSY VA model is equivalent to 
the spontaneously broken $N = 2$ LSUSY interacting theory 
containing the Yukawa interactions and mass terms, 
which is renormalizable. 

The NLSUSY VA model \cite{VA} for arbitrary $N$ \cite{BV} 
is realized by introducing (Majorana) NG fermions $\psi^i$. 
\footnote{
Minkowski spacetime indices in $D = 2$ are denoted by $a, b, \cdots = 0, 1$ 
and SO(N) internal indices are $i, j, \cdots = 1, 2, \cdots, N$. 
The Minkowski spacetime metric is 
${1 \over 2}\{ \gamma^a, \gamma^b \} = \eta^{ab} = {\rm diag}(+, -)$. 
We shall follow the convention of $\gamma$ matrices in $D = 2$ of Ref.\cite{STlin}. 
}
NLSUSY transformations of $\psi^i$ 
parametrized by constant (Majorana) spinor parameters $\zeta^i$, 
which correspond to the supertranslations of $\psi^i$ 
and Minkowski coordinates $x^a$, are given by 
\be
\delta \psi^i = {1 \over \kappa} \zeta^i 
- i \kappa \bar\zeta^j \gamma^a \psi^j \partial_a \psi^i, 
\label{NLSUSY}
\ee
where $\kappa$ is a constant whose dimension is $({\rm mass})^{-1}$. 
Eq.(\ref{NLSUSY}) satisfies a closed off-shell commutator algebra, 
\be
[ \delta_{Q1}, \delta_{Q2} ] = \delta_P(\Xi^a), 
\label{D2Ncom}
\ee
where $\delta_P(\Xi^a)$ means a translation with a generator 
$\Xi^a = 2 i \bar\zeta_1^i \gamma^a \zeta_2^i$. 

From a NLSUSY invariant differential one-form under Eq.(\ref{NLSUSY}), 
\ba
\omega^a \A \A 
= d x^a - i \kappa^2 \bar\psi^i \gamma^a d \psi^i 
\nonu
\A \A = (\delta^a_b 
- i \kappa^2 \bar\psi^i \gamma^a \partial_b \psi^i) \ dx^b 
\nonu
\A \A = (\delta^a_b + t^a{}_b) \ dx^b 
\nonu
\A \A = w^a{}_b \ dx^b, 
\label{one-form}
\ea
a NLSUSY action is obtained as the volume form in $D = 2$, 
\ba
S_{\rm NL} = \A \A - {1 \over {2 \kappa^2}} \int \omega^0 \wedge \omega^1 
\nonu
= \A \A - {1 \over {2 \kappa^2}} \int d^2 x \ \vert w \vert 
\nonu
= \A \A - {1 \over {2 \kappa^2}} \int d^2 x 
\left\{ 1 + t^a{}_a + {1 \over 2!}(t^a{}_a t^b{}_b - t^a{}_b t^b{}_a) 
\right\} 
\nonu
= \A \A - {1 \over {2 \kappa^2}} \int d^2 x 
\left\{ 1 - i \kappa^2 \bar\psi^i \!\!\not\!\partial \psi^i 
- {1 \over 2} \kappa^4 
( \bar\psi^i \!\!\not\!\partial \psi^i \bar\psi^j \!\!\not\!\partial \psi^j 
- \bar\psi^i \gamma^a \partial_b \psi^i \bar\psi^j \gamma^b \partial_a \psi^j ) 
\right\} 
\nonu
= \A \A - {1 \over {2 \kappa^2}} \int d^2 x 
\left\{ 1 - i \kappa^2 \bar\psi^i \!\!\not\!\partial \psi^i 
\right. 
\nonu
\A \A 
\left. 
- {1 \over 2} \kappa^4 \epsilon^{ab} 
( \bar\psi^i \psi^j \partial_a \bar\psi^i \gamma_5 \partial_b \psi^j 
+ \bar\psi^i \gamma_5 \psi^j \partial_a \bar\psi^i \partial_b \psi^j ) 
\right\}, 
\label{NLSUSYact}
\ea
where the second term, 
$-{1 \over {2 \kappa^2}} t^a{}_a = {i \over 2} 
\bar\psi^i \!\!\not\!\partial \psi^i$, 
is the kinetic term for $\psi^i$. 

On the other hand, we introduce $D = 2$ SO(2) vector supermultiplet 
with the SSB which is (algebraically) 
equivalent to the NLSUSY VA model for $N = 2$. 
Components fields of the off-shell L supermultiplet 
are denoted by $v^a$ for a vector field, 
$\lambda^i$ for doublet (Majorana) fermions 
and $A$ for a scalar field 
in addition to $\phi$ for a scalar field 
and $F$ for an auxiliary scalar field. 
Note that the off-shell fermionic and bosonic degrees of freedom 
of these component fields are balanced as 4 = 4. 
$N = 2$ LSUSY transformations of the component fields 
generated by $\zeta^i$ are defined as 
\ba
\A \A 
\delta_\zeta v^a = - i \epsilon^{ij} \bar\zeta^i \gamma^a \lambda^j, 
\label{LSUSY-v}
\\
\A \A 
\delta_\zeta \lambda^i 
= ( F - i \!\!\not\!\partial A ) \zeta^i 
+ {1 \over 2} \epsilon^{ab} \epsilon^{ij} F_{ab} \gamma_5 \zeta^j 
- i \epsilon^{ij} \gamma_5 \!\!\not\!\partial \phi \zeta^j, 
\\
\A \A 
\delta_\zeta A = \bar\zeta^i \lambda^i, 
\\
\A \A 
\delta_\zeta \phi = - \epsilon^{ij} \bar\zeta^i \gamma_5 \lambda^j, 
\\
\A \A 
\delta_\zeta F = - i \bar\zeta^i \!\!\not\!\partial \lambda^i. 
\label{LSUSY-F}
\ea
where $F_{ab} = \partial_a v_b - \partial_b v_a$. 
Eqs. from (\ref{LSUSY-v}) to (\ref{LSUSY-F}) 
satify the closed off-shell commutator algebra 
with a U(1) gauge transformation of $v^a$, 
\be
[ \delta_{Q1}, \delta_{Q2} ] = \delta_P(\Xi^a) + \delta_g(\theta), 
\label{D2N2com-gauge}
\ee
where $\delta_g(\theta)$ is the U(1) gauge transformation 
with a generator $\theta = - 2 (i \bar\zeta_1^i \gamma^a \zeta_2^i v_a 
- \epsilon^{ij} \bar\zeta_1^i \zeta_2^j A 
- \bar\zeta_1^i \gamma_5 \zeta_2^i \phi)$. 
A free LSUSY action which is invariant 
under Eqs. from (\ref{LSUSY-v}) to (\ref{LSUSY-F}) 
is written as 
\be
S_0 = \int d^2 x \left\{ 
- {1 \over 4} (F_{ab})^2 
+ {i \over 2} \bar\lambda^i \!\!\not\!\partial \lambda^i 
+ {1 \over 2} (\partial_a A)^2 
+ {1 \over 2} (\partial_a \phi)^2 
+ {1 \over 2} F^2 
- {\xi \over \kappa} F 
\right\}, 
\label{LSUSYact}
\ee
where the last term proportional to $\kappa^{-1}$ 
is an analog of the FI $D$ term 
with a real parameter $\xi$ satisfying $\xi^2 = 1$ 
and the field equation for the auxiliary fields 
gives the vev $<F> = {\xi \over \kappa}$ indicating the SSB. 

It can be shown that the above $D = 2$ SO(2) vector supermultiplet 
with the SSB is indeed equivalent to the $N = 2$ NLSUSY VA model 
by means of the $N = 2$ SUSY and SO(2) invariant expressions 
of $(v^a, \lambda^i, A, \phi, F)$ as composites of $\psi^i$ 
in all orders, 
\ba
\A \A 
v^a = - {i \over 2} \xi \kappa \epsilon^{ij} 
\bar\psi^i \gamma^a \psi^j \vert w \vert, 
\label{expand-v}
\\
\A \A 
\lambda^i = \xi \left[ \psi^i \vert w \vert 
- {i \over 2} \kappa^2 \partial_a 
\{ \gamma^a \psi^i \bar\psi^j \psi^j 
(1 - i \kappa^2 \bar\psi^k \!\!\not\!\partial \psi^k) \} \right], 
\\
\A \A 
A = {1 \over 2} \xi \kappa \bar\psi^i \psi^i \vert w \vert, 
\\
\A \A 
\phi = - {1 \over 2} \xi \kappa \epsilon^{ij} \bar\psi^i \gamma_5 \psi^j 
\vert w \vert, 
\\
\A \A 
F = {\xi \over \kappa} \vert w \vert 
- {1 \over 8} \xi \kappa^3 
\Box ( \bar\psi^i \psi^i \bar\psi^j \psi^j ). 
\label{expand-F}
\ea
Eqs. from (\ref{expand-v}) to (\ref{expand-F}) 
are obtained by using heuristic arguments \cite{R,STT2,STlin} 
such that the NLSUSY transformations (\ref{NLSUSY}) 
of those expressions reproduce the LSUSY ones (\ref{LSUSY-v}) 
to (\ref{LSUSY-F}). 
In particular, the transformation of Eq.(\ref{expand-v}) 
under Eq.(\ref{NLSUSY}) gives the U(1) gauge transformation 
besides the LSUSY one as 
\be
\delta_\zeta v^a(\psi) 
= - i \epsilon^{ij} \bar\zeta^i \gamma^a \lambda^j(\psi) + \partial^a X(\zeta; \psi), 
\label{NLSUSY-v}
\ee
where $X(\zeta; \psi)$ is the U(1) gauge transformation parameter 
defined by 
\be
X(\zeta; \psi) 
= \xi \kappa^2 \epsilon^{ij} \bar\zeta^i \psi^j \bar\psi^k \psi^k 
(1 - i \kappa^2 \bar\psi^l \!\!\not\!\partial \psi^l ). 
\label{gauge-parameter}
\ee
Eq.(\ref{NLSUSY-v}) means that gauge invariant quantities 
like $F_{ab}(\psi)$ transform exactly same as the LSUSY transformation. 
Since Eq.(\ref{gauge-parameter}) satisfies 
\be
\delta_{\zeta_2} X(\zeta_1; \psi) 
- \delta_{\zeta_2} X(\zeta_1; \psi) 
= - \theta(\zeta_1, \zeta_2; A(\psi), v^a(\psi), \phi(\psi)), 
\ee
the commutator on $v^a(\psi)$ of Eq.(\ref{NLSUSY-v}) 
does not contain the U(1) gauge transformation term $\delta_g(\theta)$, 
which reflects the case of the commutator on $\psi^i$ \cite{STT2,STlin}. 

By substituting Eqs. from (\ref{expand-v}) to (\ref{expand-F}) 
into the action $S_0$ of Eq.(\ref{LSUSYact}), 
it reduces to the NLSUSY VA action (\ref{NLSUSYact}) for $N = 2$ 
up to surface terms, i.e. we can show 
\be
S_0 = S_{\rm NL} + [ \ {\rm surface\ terms} \ ] 
\label{S0-SNL}
\ee
in all orders of $\psi^i$. 
Note that SO(2) is not broken in the (almost) free action. 

Let us discuss Yukawa interactions and mass terms in the $D = 2$ SO(2) vector supermultiplet 
which are invariant under Eqs. from (\ref{LSUSY-v}) to (\ref{LSUSY-F}) 
from the viewpoint of NLSUSY. 
The most general Yukawa-type interaction terms are given by 
\ba
S_f \A = \A \int d^2 x 
\{ \ f ( A \bar\lambda^i \lambda^i + \epsilon^{ij} \phi \bar\lambda^i \gamma_5 \lambda^j 
+ A^2 F - \phi^2 F - \epsilon^{ab} A \phi F_{ab} ) \ \} 
\label{Yukawa}
\ea
with $f$ being a constant whose dimension is $({\rm mass})^1$, 
while the (Majorana spinor) mass terms are 
\be
S_m = \int d^2 x 
\left\{ \ - {1 \over 2} m \ 
( \bar\lambda^i \lambda^i - 2 A F + \epsilon^{ab} \phi F_{ab} ) \ \right\}. 
\label{mass}
\ee
By substituting Eqs. from (\ref{expand-v}) to (\ref{expand-F}) 
into the terms $S_f$ of Eq.(\ref{Yukawa}) or $S_m$ of Eq.(\ref{mass}), 
surprisingly we find that $S_f$ and $S_m$ vanish respectively 
by means of miraculous cancellations among the terms, 
i.e. we can see 
\be
S_f(\psi) \equiv 0, \ \ \ S_m(\psi) \equiv 0, 
\label{SfSmvanish}
\ee
in all orders of $\psi^i$. 
This means that the most general renormalizable LSUSY action 
with the Yukawa couplings and mass terms 
of $D = 2$ SO(2) vector supermultiplet 
is equivalent to the $N = 2$ NLSUSY VA action 
up to surface terms, i.e. 
\be
S_{\rm L} \equiv S_0 + S_f + S_m = S_{\rm NL} + [ \ {\rm surface\ terms} \ ] 
\ee
in all orders of $\psi^i$. 
This new results are the breakthrough towards the realistic model building 
for the low energy particle physics and the cosmology 
in the SGM scenario \cite{ST2}. 

We summarize our results as follows. 
We have studied Yukawa-type interactions and mass terms 
in $D = 2$ SO(2) (U(1)) vector supermultiplet from the viewpoint of NLSUSY. 
We have linearized the $N = 2$ NLSUSY of VA in $D = 2$ 
by constructing the SUSY and SO(2) invariant relations 
in all orders of NG fermions 
as in Eqs. from (\ref{expand-v}) to (\ref{expand-F}). 
In the similar discussions in Refs.\cite{STT2,STlin}, 
the commutator for $v^a(\psi)$ of Eq.(\ref{expand-v}) 
are calculated by using the NLSUSY transformations (\ref{NLSUSY}) of $\psi^i$, 
and it has been shown that it does not contain the U(1) gauge transformation term. 
By using the SUSY (and SO(2)) invariant relations, 
we have proved to all orders the equivalence of the NLSUSY VA action (\ref{NLSUSYact}) 
for $N = 2$ to the free LSUSY action (\ref{LSUSYact}) up to the surface terms 
as in Eq.(\ref{S0-SNL}), 
while we have shown explicitly that the LSUSY Yukawa interactions and mass terms vanish 
respectively in Eq.(\ref{SfSmvanish}) 
by means of the cancellations among the terms. 
From these discussions we have concluded in $D = 2$ 
that the $N = 2$ NLSUSY VA model is equivalent to 
the spontaneously broken $N = 2$ LSUSY interacting theory 
containing the Yukawa interactions and mass terms, 
which is renormalizable. 
From the sytematics in the cancellations in $D = 2$ 
we anticipate that this is also the case in $D = 4$. 
In fact, we have observed the cancellations 
among mass terms at the leading order in $D = 4$ $N = 1$. 
As for the all-order cancellations in $D = 4$, we need futher calculations. 

Finally we just mention the physical meaning of our results. 
It is a breakthrough towards the realistic unified model building 
based upon the compositeness of all particles (except graviton), 
a la Landau-Gintzburg theory 
v.s. Bardeen-Cooper-Schriefer theory for the superconductivity, 
though non-relativistic, which is the prototype of SGM scenario, 
i.e. SM (GUTs) v.s. SGM (NLSUSY GR) as mentioned in \cite{KS}. 
Because NLSUSY VA model is the simplest and unique NLSUSY realization 
and is accommodated as the cosmological term for everything 
(except the spacetime curvature energy) in the SGM scenario. 
The equivalence between the unique NLSUSY VA model 
and the various LSUSY free action with FI terms are well understood. 
The generic nontrivial Yukawa and the mass terms 
break the equivalence and are excluded. 
In our work we have found that these terms 
become trivial (vanish) automatically 
only by the composite viewpoint (i.e. the SUSY invariant relations), 
therefore they can be added to the equivalent free action 
without violating the equivalence. 
Once they are added to the free action with FI term 
(i.e. thrown into the true vacuum), 
they become nontrivial, break SUSY and produce mass 
for the composite fields automatically as partly realized in \cite{ST2}.

%
%

\newpage

%
\newcommand{\NP}[1]{{\it Nucl.\ Phys.\ }{\bf #1}}
\newcommand{\PL}[1]{{\it Phys.\ Lett.\ }{\bf #1}}
\newcommand{\CMP}[1]{{\it Commun.\ Math.\ Phys.\ }{\bf #1}}
\newcommand{\MPL}[1]{{\it Mod.\ Phys.\ Lett.\ }{\bf #1}}
\newcommand{\IJMP}[1]{{\it Int.\ J. Mod.\ Phys.\ }{\bf #1}}
\newcommand{\PR}[1]{{\it Phys.\ Rev.\ }{\bf #1}}
\newcommand{\PRL}[1]{{\it Phys.\ Rev.\ Lett.\ }{\bf #1}}
\newcommand{\PTP}[1]{{\it Prog.\ Theor.\ Phys.\ }{\bf #1}}
\newcommand{\PTPS}[1]{{\it Prog.\ Theor.\ Phys.\ Suppl.\ }{\bf #1}}
\newcommand{\AP}[1]{{\it Ann.\ Phys.\ }{\bf #1}}

\end{document}